\documentclass[a4paper,11pt]{article}
\usepackage[utf8]{inputenc}
\usepackage[english]{babel}
\usepackage{amsfonts}     
\usepackage[psamsfonts]{amssymb}
\usepackage[bottom]{footmisc}
\usepackage{amsmath}
\usepackage{graphicx}
\usepackage[dvipsnames]{xcolor}
\usepackage[final]{pdfpages}

\allowdisplaybreaks                    
\def\cH{{\cal{H}}}

\def\cA{{\cal{A}}}
\def\cM{{\cal{M}}}
\def\cW{{\cal{W}}}

\def\exterior{{{\raise0.2em\hbox{$\scriptstyle\bigwedge$}}{}}}

\def\ket#1{\vert #1 \rangle}            
\def\p{\partial}

\def\dfrac #1#2{\displaystyle{\frac{#1}{#2}}}

\begin{document}

\title{Laplacian on  fuzzy de~Sitter space}

\author{Bojana Brki\' c, Maja Buri\'c and Du\v sko Latas\thanks{bojana.brkic@ff.bg.ac.rs, majab@ipb.ac.rs, latas@ipb.ac.rs}
                   \\[15pt]
        University of Belgrade,  Faculty of Physics, Studentski trg 12
                   \\
        SR-11001 Belgrade
       }

\maketitle

\begin{abstract}
We study details of geometry of  noncommutative  de~Sitter space: we determine the Riemann and Ricci curvature tensors, the energy and the Laplacian. We find, in particular, that fuzzy de~Sitter space is an Einstein space, $R_{ab}=-3\zeta\,\eta_{ab}$.  The Laplacian, defined in the noncommutative frame formalism, is not hermitian and gives nonunitary evolution. When symmetrically ordered, it has the usual quadratic form $\,\Delta=\Pi_a\Pi^a$ (when acting on functions in  representation space, $\Psi\in\cH$): we find its eigenstates and discuss its spectrum. This result is a first step in a study of the scalar field Laplacian, $\, \Delta = [\Pi_a, [\Pi^a,\ \,]]\,$,  and its propagator.
\end{abstract}

\section{Introduction}

The relevance of noncommutative or fuzzy spaces is, both in mathematics and theoretical  physics, many-fold. Mathematically, the concept is interesting as it provides an extension of geometry to algebraic structures like $C^*$ or matrix algebras: in particular, it is  important to generalize the notion of smoothness to discrete structures such as algebras of matrices. Physically, discreteness of observables is usually related to their quantization: it is commonly expected that  spacetime is quantized at the Planck scale, i.e. that coordinates are represented by  operators. If we  describe fields as functions of spacetime variables (that is, as elements of the algebra of coordinates, $\cA$), then in order to write their equations of motion we need to define Laplace and Dirac operators. Further, if additional differential-geometric quantities like connection and curvature are defined, we have achieved a description of gravity at the  Planck scale, or more precisely, at the noncommutativity scale.

An alternative way to obtain quantum spacetime is to express coordinates, metric etc. through a set of fields which are assumed to be elementary, for example strings. Then, if the fundamental structure is a linear (quantum field) theory,  the effective geometric structure will likely be linear too, presumably described by operators. It is perhaps in such case less natural to expect that quantum spacetime has a differential-geometric structure. However, if the classical limit exists, effective quantities corresponding to metric, connection and curvature exist as well, with their usual geometric properties.

 Fuzzy de Sitter space  is a noncommutative space defined in the framework of the  noncommutative frame formalism \cite{book}, using the Lie-algebra structure of the de~Sitter group $SO(1,4)$.  The formalism gives a general definition  of noncommutative differential geometry  adjusted to gravity, and thus has a potential to describe gravity (or effective gravity) as geometry, in the quantum regime. However, it had been mostly  applied to  lower-dimensional models, while the real advance would be to find noncommutative versions of physical solutions to Einstein equations like the Schwarzschild black hole or the FLRW cosmologies (with unbroken spherical symmetry and beyond the tensor-product constructions). 

One of distinctive properties of the noncommutative frame formalism is that it singles out a set of derivations $e_a$ and the corresponding dual  1-forms $\theta^a$ that define the free-falling frame. Metric $  g\, $ is a linear function with constant values in the frame basis,
\begin{equation}
 g^{ab}=g(\theta^a\otimes\theta^b) = \eta^{ab} .
\end{equation}
A sufficient condition that the metric components be constant is
\begin{equation}
 [f,\theta^a]=0   ,                     \label{theta}
\end{equation}
where $f$ is an arbitrary function of noncommutative coordinates, $f\in\cA$: this condition is a part of  definition of the moving frame. Frame derivations $e_a $ are generated by  momenta $\Pi_a$\footnote{We replace here the usual notation for the operators of momenta $p_a$ by $\Pi_a$ to avoid possible confusion, as in the representations we are using wave functions $\Psi$  are functions of the momentum variables, $\Psi(p_0, \vec p)$.},
\begin{equation}
 e_a f=[\Pi_a,f] ,                           \label{via}
\end{equation}
and $\Pi_a$ are usually taken to be antihermitian elements of $\cA$. From these initial assumptions one can build a differential geometry in close analogy to the commutative one.

In this paper we investigate geometric properties of fuzzy de Sitter space additional to those already discussed in \cite{Buric:2015wta,Buric:2017yes}; immediate consequences of fuzzy de~Sitter geometry to cosmology were previously analyzed in \cite{Buric:2017yes,Buric:2019yau}. After reviewing the representations of the $SO(1,4)$  which we use in Section~2,   in Section~3 we introduce the operator of energy of  fuzzy de~Sitter space and determine its spectrum. In Section~4 we calculate the Riemann and  Ricci curvature tensors and the Laplacian. In Section~5 we find the spectrum and the eigenfinctions of the quantum-mechanical Laplacian in $\,(\rho, s=0,\frac 12)\,$ unitary irreducible representations of the principal continuous series. A discussion of  the presented results, of  properties and limitations  that they give and of  possible further research directions is given in the concluding section.

\section{Review of the representation}

Fuzzy de Sitter space is defined using the  Lie algebra of the de~Sitter group $SO(1,4)$. Its generators $\cM_{\alpha\beta}$ satisfy
\begin{equation}
 [\cM_{\alpha\beta}, \cM_{\gamma\delta}] = - i(\eta_{\alpha\gamma} \cM_{\beta\delta}
 -\eta_{\alpha\delta} \cM_{\beta\gamma}-\eta_{\beta\gamma}
 \cM_{\alpha\delta}+  \eta_{\beta\delta} \cM_{\alpha\gamma})   ,      \label{SO14}
\end{equation}
$\alpha,  \beta = 0,1,2,3,4\,$ and the  signature is  $\, \eta_{\alpha\beta}\,$= diag$(1,-1,-1,-1,-1)$. In  unitary representations $\cM_{\alpha\beta}$ are hermitian. Vector $\cW^\alpha$, a generalization of the Pauli-Lubanski vector, is defined by
\begin{equation}
 \cW^\alpha =\dfrac 18 \,\epsilon^{\alpha\beta\gamma\delta\eta}
 \cM_{\beta\gamma} \cM_{\delta\eta}         \,    .  \label{W}
\end{equation}
The  $SO(1,4)$ has a quadratic and a quartic Casimir operator, 
\begin{equation}
C_2=  {\cal Q} = -\frac 12 \, \cM_{\alpha\beta} \cM^{\alpha\beta}\, ,\qquad C_4=
{\cal W} =-\eta_{\alpha\beta}\, \cW^\alpha \cW^\beta  .
\end{equation}
In unitary irreducible representations (UIR) of the $SO(1,4)$,   labelled by  quantum numbers $(\rho, s)$,  values of the Casimir operators are
\begin{equation}
 {\cal Q} = -s(s+1) +\frac 94 +\rho^2 ,\qquad {\cal W} = s(s+1)\Big(\frac 14 +\rho^2\Big) .\label{Casimirs}
\end{equation}
There are three series of unitary irreducible representations:  the principal continuous series with $\,\rho\in{\mathbb R}$, $\rho\geq 0$, $s=0,\frac 12, 1,\dots $, \, the  complementary continuous series, $\,i\rho=\nu\in{\mathbb R}$, $\,\vert\nu\vert<\frac 32$, $\, s=0, 1, 2,\dots$, and  two discrete series, $\,\frac 12 +i\rho=q$, $\, q =s,s-1,\dots 0\,$ or $\frac 12\, $, \ $s=\frac 12, 1,\frac 32,\dots\ $. 

In analogy with the fuzzy sphere \cite{fs}, fuzzy de Sitter space can be defined as a `noncommutative embedding' or  hypersurface in  five-dimensional noncommutative space. The latter is generated by coordinates 
\begin{equation}
 x^\alpha = \ell W^\alpha ,                    \label{coordinates}
\end{equation}
while the embedding  is realized through the  second Casimir relation,
\begin{equation}
\eta_{\alpha\beta}\, x^\alpha x^\beta = -\ell^2 {\cal W} = -\frac{3}{\Lambda} \ ,
\end{equation}
 where $\Lambda$ is the cosmological constant. As in the case of fuzzy Anti-de~Sitter space 
  \cite{Jurman:2013ota}, not every UIR gives a physically meaningful realization of  de~Sitter space (assuming the identification  (\ref{coordinates})). For example, (\ref{Casimirs}) implies that $\Lambda$ is positive just in: all  representations of the principal continuous series, representations of complementary  series with $\vert \nu\vert <\frac 12$, and  representations of discrete series with $\,q=s=0,\,\frac 12\,$. Therefore,  discreteness of the cosmological constant (which appears naturally in discrete series) is in fact not generic. We will, in the following,  discuss properties of fuzzy de~Sitter space defined as principal continuous series representations $(\rho,s)$:  concrete calculations are done for values $s=0\,$ and $\,s=\frac 12\,$. 

The linear space of vector fields ${\cal X}(\cA)$  on a noncommutative algebra $\cA$ is infinite-dimensional. In order to restrict  dimension of the tangent space $T(\cA)$ associated to $\cA$, it is usual to define it as a subspace $T(\cA) \subset {\cal X}(\cA) $ by giving it a finite basis, for example $\{ e_a\}$ defined by  (\ref{via}). There are two sets of momenta $ \Pi_a$ that give, in  application of the frame formalism to fuzzy de~Sitter space,  the de Sitter metric. One  is 
\begin{equation}
  i\Pi_a = \sqrt{\zeta\Lambda}\,\cM_{\alpha\beta}   \,      ,         \label{pa,first}
\end{equation}
where $\zeta\,$ is a numerical factor fixed by the value of the curvature scalar and index $a$ denotes  antisymmetric pairs $\{\alpha\beta\}$. In this case  we obtain the metric with coordinate components equal to\footnote{In the commutative limit $g^{\alpha\beta}$  projects to the de~Sitter hypersurface.} 
\begin{equation}
 g^{\alpha\beta} =e^\alpha_a e^\beta_b\eta^{ab} =3\eta^{\alpha\beta} -\Lambda x^\beta x^\alpha\, .
\end{equation}
The  tangent  space is ten-dimensional. Second, `compact' choice which gives a four-dimensional tangent space is to define momenta as operators proportional to dilation and translations, \cite{Buric:2015wta}
\begin{equation}
 i\Pi_0= \sqrt{\zeta\Lambda}\, \cM_{04} ,\qquad i\Pi_i= \sqrt{\zeta\Lambda}\,(\cM_{i4}+ \cM_{0i})\, , \quad i=1,2,3\, .             
 \label{momenta}
\end{equation}
It implies the line element
\begin{equation}
 ds^2 = d\tau^2 -e^{2\tau}\,(dx^i)^2 \,,
\end{equation}
 with cosmic time $\tau$ and conformal time $\eta $  given by\footnote{The given sign of $\eta$  correspods to the upper half of the de Sitter manifold, \cite{BD}.}
\begin{equation}
 \tau = \ell\, \log(\cW_0-\cW_4)\, , \qquad
  \eta = -\ell\, e^{-\tau/\ell} = \cW_0-\cW_4 .                       \label{time}
\end{equation}

 Apparently, the most viable Hilbert space representation of the principal continuous series and the one that we used in calculations was  given by Moylan, \cite{Moylan}. It is defined on the space of  UIR's of the Poincar\' e group of mass $m$ and spin $s$, \cite{Bargmann}: the corresponding Hilbert space  is a direct sum of two Hilbert spaces of states with positive and negative energies, $\cH(m,s,+)\oplus \cH(m,s,-)\,  $. The group generators, denoted  by $\, \widetilde{\cM}_{\alpha\beta}$, are 
\begin{eqnarray}
&&\  \widetilde{ \cM}_{\alpha\beta} = \begin{pmatrix}
                      M_{\alpha\beta} & 0\\[4pt]
                      0 & M_{\alpha\beta}
                     \end{pmatrix}           ,                     \label{tilde}   \\[4pt]
&& \begin{array}{ll}
   M_{ij} = i\Big(p_i \, \dfrac{\p}{\p p^j}-p_j\, \dfrac{\p}{\p p_i}\Big) + S_{ij},  &  M_{i0} = - i p_0\,\dfrac{\p}{\p p^i} + S_{i0},\\[16pt]
   M_{4j} = -\dfrac \rho m \, p_j -\dfrac{1}{2m} \,\{p^0, M_{0j}\}-\dfrac{1}{2m} \,\{p^i, M_{ij}\} ,\quad
   & M_{40} = -\dfrac \rho m \, p_0 +\dfrac{1}{2m} \,\{p^i, M_{0i}\} .\qquad
 \end{array}         \label{13}
\end{eqnarray}
To simplify,  we will in the following rescale $\,p_\mu\,$ to be dimensionless, $\, p_\mu/m \to p_\mu$. The  $S_{\mu\nu}$  in (\ref{13}), $\mu, \nu = 0,1,2,3$, are the usual spin generators:  for the scalar UIR $\,(\rho,s=0)$,  $\, S_{\mu\nu}=0\,$;  for the spinor representation $\,(\rho,s=\frac 12)\,$, $\, S_{\mu\nu} = \frac i4 [\gamma_\mu,\gamma_\nu]$. 

Though the Moylan Hilbert-space representation is, at the level of the algebra, block-diagonal, it is not reducible: this is a very fine point which shows up (as noted and commented in \cite{Moylan})  in the non-self-adjointness of  operators $\, M_{\alpha\beta}$ that  appear diagonally in (\ref{tilde})\footnote{We thank Ilija Buri\' c for discussions of the G\aa rding domain which brought this fact into our focus: in the previous paper \cite{Buric:2019yau} we erroneously assumed that the only consequence of the direct sum is  doubling of states.}.   However, the $\,\widetilde{\cM}_{\alpha\beta}\,$  are hermitian in  $\,\cH(m,s,+)\oplus \cH(m,s,-)\, $. This can be best  seen by mapping the given representation to an equivalent one, defined in the direct sum of two positive-energy Hilbert spaces,
\begin{equation}
 \theta:\, \cH(m,s,+)\oplus \cH(m,s,-)\to \cH(m,s,+)\oplus \cH(m,s,+) \equiv \cH_\uparrow \oplus \cH_\downarrow \ .
\end{equation}
We will use the latter  in the calculations; thus we denote
 \begin{equation}
   \theta\,\widetilde\Psi \equiv\Psi \, , \qquad
      \theta \widetilde{\cM}_{\mu\nu}\theta^{-1} \equiv \cM_{\mu\nu}\, ,
 \end{equation}
 for $\,\widetilde\Psi \in \cH(m,s,+)\oplus \cH(m,s,-)\, $ and $\,\Psi \in \cH(m,s,+)\oplus \cH(m,s,+)\, $. We shall however not discuss in details the difference in hermiticity properties of $\,M_{\alpha\beta}\,$ and   $\,\cM_{\alpha\beta}\,$: we only comment the case  of spherically symmetric operators, where the transition from $\,\cH(m,s,+) $  to $\,\cH(m,s,+)\oplus \cH(m,s,+)  $  reduces to the change of interval of radial variable from $\,z\in (0,1)\,$ to $\,z\in (0,\infty)\,$.  This change affects  hermiticity of the corresponding operators in a clear way, through the boundary conditions imposed on functions generating the Hilbert space.
 
 The action of $\, \theta $ is defined by, \cite{Moylan}
 \begin{eqnarray}
  \theta\begin{pmatrix}
                            \ket{ \vec p\, s_3\,+}\\[4pt]
                              \ket{ \vec p\, s_3\,-} 
                           \end{pmatrix}= \begin{pmatrix}
                            \ket{ \vec p\, s_3\,+}\\[4pt]
                              \ket{- \vec p\, s_3\,+} 
                          \end{pmatrix},
                          \qquad
      \theta\begin{pmatrix}
                      p_\mu & 0\\[4pt]
                      0 & p_\mu
                     \end{pmatrix}\theta^{-1}= \begin{pmatrix}
                      p_\mu & 0\\[4pt]
                      0 & -p_\mu
                     \end{pmatrix},                           
\label{block}
\end{eqnarray}
and implies
\begin{equation}
 \cM_{\mu\nu}=\begin{pmatrix}
                      M_{\mu\nu} & 0\\[4pt]
                      0 & M_{\mu\nu}
                     \end{pmatrix} , \qquad 
        \cM_{\mu 4}=\begin{pmatrix}
                      M_{\mu 4} & 0\\[4pt]
                      0 & - M_{\mu 4}
                     \end{pmatrix} .
                  \label{generators}
\end{equation}

The scalar product  is  invariant to $\theta\,$ and positive-definite.  However,  due to specific form of (\ref{block}), it acquires a somewhat counter-intuitive form after the $\theta$-mapping. Introducing 
\begin{equation}
 \Psi =\begin{pmatrix}  \psi_\uparrow \\ \psi_\downarrow
                                           \end{pmatrix}          \label{wave}
\end{equation}                                        
 the scalar product can be expressed as
 \begin{equation}
  (\Psi, \Psi') = (\psi_\uparrow,\psi_\uparrow ') + (-1)^{2s} (\psi_\downarrow,\psi_\downarrow ') ,
 \end{equation}
where $\,(\psi,\psi')\,$ in the last formula depends on the spin: for  cases we discuss we have, \cite{Moylan, Bargmann} 
\begin{eqnarray}
&& (\psi, \psi') = \int \frac{d^3p}{2\vert p_0\vert}\, \,\psi^*\psi'\, ,\qquad s=0\, ,\label{(s=0)}
\\[4pt]
&& (\psi, \psi') = \int \frac{d^3p}{2\vert p_0\vert}\, \,\psi^\dagger \gamma^0\psi'\, ,\quad s=\frac 12\ .                                                 \label{(s=1/2)}
\end{eqnarray}
  
 It is perhaps  at this point appropriate  to note that the Moylan  representation, though with a number of advantages, is in several aspects inconvenient to use: apart from the problem with hermiticity (appearing because of the splitting of the Hilbert space into a direct sum), the scalar product  is  different for each value of spin. More practical seems to be the Hilbert space representation of the principal continuous series that is commonly used in conformal field theory, \cite{Dobrev}.

\section{Energy}

Let us  discuss the spectrum of energy of fuzzy de~Sitter space. In  conformal field theory energy ${\cal E}\,$ is often identified with the dilation generator $\cM_{04}$; here we have
\begin{equation}
 [i\cM_{04}, \cW_0-\cW_4]=\cW_0-\cW_4 \, ,
\end{equation}
 that is, $\,\cM_{04}$  is canonically conjugate to the cosmic time $\tau$. Therefore we define  energy as
\begin{equation}
 {\cal E} =\frac \hbar \ell\, \cM_{04} =   \frac{i \hbar}{l\sqrt{\zeta\Lambda}}\, \Pi_0 \, .
\end{equation}

We solve the energy eigenvalue equation in representations $\,(\rho,s=0)$. In subspace $\, \cH_\uparrow\,$,     $\,\cM_{04}$ reduces to
\begin{equation}
 M_{04,\uparrow}  = M_{04} = p_0\left( \rho - \frac{3i}{2} -ip\, \frac{\p}{\p p}\right) .
\end{equation}
Here $ p$ is the radial component of  $\,\vec p\,$,  $\, p^2=\vec p^2 = -p_i p^i\,$, and $p_0$  denotes the positive square root, $\, p_0 = \sqrt{p^2+1}$. The $\,M_{04}$ commutes with the angular momentum  $\,M_{ij}$, thus the angular variables  can be separated in  equation
\begin{equation}
 M_{04,\uparrow} \, \psi_\uparrow= \lambda \,\psi_\uparrow .
\end{equation}
 Using the Ansatz
\begin{equation}
 \psi_{\lambda l m,\uparrow}(\vec p) = \psi_{\lambda,\uparrow}(p)\, Y_l^m(\theta, \varphi) =\frac{f_{\lambda,\uparrow}(p)}{p}\,Y_l^m(\theta, \varphi)  \, ,
\end{equation}
we obtain the  radial equation
\begin{equation}
 i\big(p_0^2-1\big) \, \frac{d \psi_{\lambda,\uparrow}}{dp_0} +\Big( \frac{3i}{2} - \rho\Big) p_0 \, \psi_{\lambda,\uparrow}=-\lambda \, \psi_{\lambda,\uparrow} \, .
\end{equation}
It has a solution  
\begin{equation}
 \psi_{\lambda,\uparrow} = c_\lambda\, p^{-\frac 32- i\rho}\, \Big( \frac{p_0-1}{p_0+1} \Big)^{\frac{i\lambda}{2}}\,  
                        \label{solutionE}
\end{equation}
or, written in variable $\, z$ defined in  (\ref{z}), 
\begin{equation}
 f_{\lambda,\uparrow} =  
                       C_\lambda\, (1-z^2)^{\frac 12+i\rho } \, z^{-\frac 12-i\rho+i\lambda}\ .  \label{solE}
\end{equation}
The radial equation is the same in  subspace $\,\cH_\downarrow$, with  $p_0$ replaced by $-p_0$ or  $\lambda$  by $-\lambda$; therefore
\begin{equation}
  f_{\lambda,\downarrow}(p) = f_{-\lambda,\uparrow}(p)\, .
\end{equation}

The radial solutions  behave, upon integration, as plane waves in   $\,\log z$. The eigenvalue $\lambda\,$ is not restricted: $\,\lambda \in \mathbb{R}\,$, and $\,{\cal E}$ has  continuous spectrum. It is perhaps instructive to check explicitly the norm of the eigenfunctions. From (\ref{(s=0)}), for solutions (\ref{solutionE}) we obtain, \cite{skripta}
\begin{equation}
( \psi_{\lambda,\uparrow}, \psi_{\lambda',\uparrow}) = \delta_{ll'}\,\delta_{mm'}\, C_\lambda^* C_{\lambda '} \int\limits_0^1 \frac{dz}{z}\, z^{i\,(\lambda'-\lambda)}= 
\delta_{ll'}\,\delta_{mm'}\, C_\lambda^* C_{\lambda '} \, \left\{ 
\begin{array}{ll}
                                          \pi \, \delta (\lambda'-\lambda),\quad &  \lambda'=\lambda    \\[6pt]
                                           \frac{1}{\, i(\lambda'-\lambda)}\, , & \lambda' \neq \lambda   \end{array} 
\right. .
\end{equation}
In fact, the eigenfunctions are properly normalized  only in the full Hilbert space $\,\cH_\uparrow \oplus \cH_\downarrow\,$:
\begin{equation}
 ( \psi_{\lambda}, \psi_{\lambda'}) = ( \psi_{\lambda,\uparrow}, \psi_{\lambda',\uparrow}) +( \psi_{\lambda,\downarrow}, \psi_{\lambda',\downarrow}) =( \psi_{\lambda,\uparrow}, \psi_{\lambda',\uparrow}) +( \psi_{-\lambda,\downarrow}, \psi_{-\lambda',\downarrow}) = \delta_{ll'}\,\delta_{mm'}\, \delta (\lambda'-\lambda)  \nonumber
\end{equation}
for $C_\lambda=\sqrt{1/{2\pi}}$.

In the  ($\rho,s=1/2$) UIR's  the radial eigenfunctions have a slightly different form because the measure in the scalar product differs;  the energy spectrum and  normalization are the same.

\section{Laplacian, general aspects}

In order to find  expression for the Laplacian on fuzzy de~Sitter space we give a few  definitions; details can be found in reference \cite{book} which is dedicated to the noncommutative frame differential calculus.

Differential geometry of fuzzy de Sitter space was partly discussed in \cite{Buric:2015wta}. It is, in the frame formalism, completely defined by the operators of momenta $ \Pi_a$ and their algebra. Intuitively speaking, momenta generate infinitesimal translations.  They need not belong to  spacetime algebra $\cA$ (indeed in particular case of commutative spaces they do not);  $ \Pi_a$ are usually taken to be antihermitian. This assumption is in \cite{book} called the `reality condition': it means that  frame derivations map real functions (i.e., hermitian operators) to real functions. 

Differential of a function $f\in \cA\,$ is defined as $\, df = (e_a f)\,\theta^a\, $. Assuming that $d^2=0$ and that condition (\ref{theta}) is consistent with the action of $d$, as well as that  $ \theta^a\, $ form a base to the cotangent space, one finds that momenta satisfy relations of the form
\begin{equation}
 2P^{ab}{}_{cd}\,  \Pi_a \Pi_b +C^a{}_{cd}\,\Pi_a +K_{cd} = 0\, ,        \label{Pabcd}
\end{equation}
where all coefficients are constant: $\Pi_a$'s cannot be chosen  arbitrarily. The action of $\,d$ can be extended to 1-forms, the exterior multiplication can be introduced. Clearly, 1-forms in general do not anti-commute, not even the  $\,\theta^a$'s. However, the structure of the  exterior algebra $\Omega^*(\cA)$ is closely related to the structure of momentum algebra  (\ref{Pabcd}). As shown in \cite{book}, a consistent way to introduce the exterior product is to define
\begin{equation}
\theta^a\wedge \theta^b \equiv\theta^a \theta^b =  P^{ab}{}_{cd} \, \theta^c\theta^d \, ,
\qquad     d\theta^a =-\frac 12\, C^a{}_{bc}\, \theta^b\theta^c .
\end{equation}

 The momentum algebra  in the fuzzy de Sitter case is a Lie subalgebra of the $so(1,4)$,
\begin{equation}
 [\Pi_0,\Pi_i]=\sqrt{\zeta\Lambda}\,\Pi_i, \qquad [\Pi_i,\Pi_j]=0 ,              \label{*}
\end{equation}
so the nonvanishing structure constants are
\begin{equation}
 C^i{}_{0j} = - C^i{}_{j0} = \sqrt{\zeta\Lambda}\, \delta^i_j  \, ,  \qquad 
 P^{ab}{}_{cd} =\frac 12\,\big(\delta^a_c\delta^b_d -\delta^a_d\delta^b_c\big) \, .
\end{equation}
The Lie-algebra structure of (\ref{*}) implies that  the frame 1-forms anticommute, $\, \theta^a \theta^b = -\theta^b \theta^a$. Furthermore, the connection 1-form $\, \omega^a{}_b = \omega^a{}_{cb} \theta^c\,$ can be taken to be  
\begin{equation}
 \omega_{abc} =\frac 12\,(C_{abc}+C_{cab}-C_{bca}) \,
\end{equation}
as in the commutative case, so we have
\begin{equation}
 \omega^0{}_0=0,\qquad \omega^0{}_j=\sqrt{\zeta\Lambda}\, \eta_{ij}\theta^i\, ,\qquad 
  \omega^i{}_0= -\sqrt{\zeta\Lambda}\, \theta^i\, ,\qquad \omega^i{}_j =0 \,.         \label{connection}
\end{equation}
This connection is metric-compatible and torsionless. For the Riemann curvature, $\,\Omega^a{}_b = \frac 12\, R^a{}_{bcd}\, \theta^c\theta^d  $,
we find
\begin{eqnarray}
&&  \Omega^0{}_0=0,\qquad \Omega^0{}_j=-\zeta\Lambda\, \eta_{ij}\theta^0\theta^i\, ,\qquad 
  \Omega^i{}_0= \zeta\Lambda\, \theta^0 \theta^i\, ,\qquad \Omega^i{}_j =-\zeta\Lambda \,\eta_{ik} \theta^i\theta^k\, ,      
\\[6pt]
&& R_{00} =-3\zeta\Lambda\, \eta_{00}\,, \qquad R_{ij}=-3\zeta\Lambda\, \eta_{ij}\,, \qquad   R=6\zeta\Lambda\,,
\end{eqnarray}
where $\, R_{bd}=R^a{}_{bad}\,$ and $\, R=R^a{}_a$. The Ricci tensor  satisfies relation   $\, R_{ab}= -3\zeta\Lambda\,\eta_{ab}\,$, that is,  fuzzy de Sitter space is a (noncommutative) Einstein manifold.

In order to define the Laplacian $\,\triangle\,$ we need the Hodge-dual $\, *$ and the co\-differential $\delta$. Noncommutative $\delta\,$ is defined as the corresponding commutative operator: its action  on  $p$-forms is $ \  \delta =(-1)^{np+n+1} *d\, *\,  ,    $ where $n$ is dimension of the cotangent  space. 
Noncommutative Hodge-dual  differs from the standard one only when the frame 1-forms do not anticommute, like e.g. on the truncated Heisenberg algebra \cite{Buric:2010xs}; here it is straightforward. The Laplacian is then given by
\begin{equation}
  \triangle = d\delta + \delta d\,      .     \label{Delta}
\end{equation}

Applying  (\ref{Delta}) to 0-forms we find the action of the Laplacian on scalar fields, i.e. scalar functions of noncommutative coordinates:
\begin{equation}
 \triangle f = [\Pi_0,[\Pi_0,f]]+[\Pi_i,[\Pi^i,f]]-3\sqrt{\zeta\Lambda}\,[\Pi_0,f] \, .      \label{Deltaf}
\end{equation}
Analogously, its action on  wave functions, elements of the representation space, is given by
\begin{equation}
\triangle\Psi = \big( \Pi_0 \Pi^0+\Pi_i \Pi^i-3\sqrt{\zeta\Lambda}\,\Pi_0 \big) \Psi \,.               \label{DeltaPsi}
\end{equation}
Neither of the two actions is  hermitian: this property is unusual and probably unwanted. In the commutative case, hermiticity of the  Laplacian is guaranteed by uniqueness of the de~Rham calculus and by invariance properties of the integral measure. In the noncommutative case differential calculus is not unique, and the integral (trace) is  defined only in concrete representation. A nonhermitian Laplacian similar to (\ref{Deltaf}) was obtained for the $h$-deformed Lobachevsky plane,  \cite{Madore_1999,Cho_1999}. The problem of hermiticity  was there solved  by changing  the operator ordering in \cite{Madore_1999}, and by changing the definition of the adjoint in \cite{Cho_1999}.  Here we  act similarly: we define the Laplacian by
\begin{equation}
 \Delta\Psi = \frac 12\,(\triangle +\triangle^\dagger)\, \Psi = (\Pi_0 \Pi^0+\Pi_i \Pi^i)\Psi  \, .           \label{deltapsi}
\end{equation}

It is perhaps worth mentioning that, for the first set of momentum operators (\ref{pa,first}) (that we do not consider), the Laplacian reduces to  quadratic Casimir operator $C_2$. This means that,  acting on the wave functions $\Psi$,  Laplacian is a constant. The spectrum of  scalar field $\,\Delta\,$  is then given by the branching rules for the tensor product of the utilized representations.

\section{Laplacian, representation}

 In  commutative curved spacetime the classical equation of motion for the scalar field $f$   is
 \begin{equation}
  (\Delta +\mu^2 +\xi R) \,f  =0           ,              \label{KG}
 \end{equation}
 where $\mu\,$ is mass of the field  and  $\, \xi\,$ is the coupling to  curvature.  Since in  de~Sitter space the scalar curvature $R$ is constant,  (\ref{KG})  has the form of the eigenvalue equation for the Laplacian,
 \begin{equation}
 (\Delta+ M^2)f =0,  \qquad     
   M^2 =\mu^2+\xi R \,  . \label{equ}
 \end{equation}
 Particular solutions to (\ref{KG}) constitute a basis to the Hilbert space of solutions $\cH\,$ that is  adapted for quantization:  positive-energy solutions define the one-particle Hilbert space of states $\cH_+$  that gives  quantum-mechanical description of the scalar particles\,\footnote{Division to positive- and negative-energy states can be done in static spacetimes. However, partitioning $\cH=\cH_+\oplus\cH_-$ (which is not unique and depends on the choice of coordinates) can be done  generally, \cite{Wald}.  }. 
 
 It is important to emphasize that in the noncommutative framework the classical equation of motion for the scalar field $\,f\in \cA$,
 \begin{equation}
 [\Pi_0,[\Pi^0,f]]+[\Pi_i,[\Pi^i,f]]+ M^2 f =0\, ,         \label{EQ}
 \end{equation}
differs from the quantum-mechanical equation for the scalar particle  described by the wave function $\,\Psi\in \cH$,
\begin{equation}
  (\Pi_0 \Pi^0+\Pi_i \Pi^i)\Psi  + M^2\Psi =0 \,      .       \label{45}
\end{equation}
 In simple cases solutions to these two equations are related.  Another important observation is that, in  fuzzy de~Sitter space, eigenstates of the Laplacian (\ref{deltapsi}) do not have  definite values of energy because the two observables are not compatible,  $\,[{\cal E},\Delta]\neq 0$. This obstructs  direct interpretation of the positive-energy subspace $\cH_+$ as space of one-particle excitations of the quantum field.

Fuzzy de~Sitter Laplacian, mapped to $\cH_\uparrow\oplus\cH_\downarrow\,$, is block-diagonal
\begin{equation}
  \Delta =  \begin{pmatrix}
                 \Delta_\uparrow & 0\\[2pt] 0 &   \Delta_\downarrow
                \end{pmatrix}  .
\end{equation}
To simplify  already cumbersome notation, we rescale in equations that follow $\, \Delta\to\zeta\Lambda\,\Delta\, $ and $\ \zeta\Lambda \,M^2 \to M^2\,$. We  solve (\ref{KG})  in $(\rho,s=\frac 12)\,$  representations. The wave functions are of the form (\ref{wave}), where  $\,\psi_{\uparrow,\downarrow}(\vec p)\,$ are  bispinors, solutions to the momentum-space Dirac equation. They are given, in accordance with \cite{Buric:2019yau}, (\ref{block}), by
\begin{equation}
 \psi_\uparrow(\vec p) = \begin{pmatrix} \varphi_\uparrow(\vec p)\\[8pt]  \dfrac{p_k\sigma^k}{1+p_0}\, \varphi_\uparrow(\vec p)
 \end{pmatrix}  ,\qquad  
             \psi_\downarrow(\vec p) = \begin{pmatrix} \varphi_\downarrow(-\vec p)\\[8pt] - \dfrac{p_k\sigma^k}{1-p_0}\,\varphi_\downarrow(-\vec p)                                                                                                                                                                                                                                                                                       \end{pmatrix}                      .        \label{bispinors}
\end{equation}
The $\,\varphi_{\uparrow,\downarrow}\,$ are spinors, and the scalar product is given by (\ref{(s=1/2)}),
 \begin{equation}
  (\Psi, \Psi') = (\psi_\uparrow,\psi_\uparrow ') -  (\psi_\downarrow,\psi_\downarrow ') 
  =\int \frac{d^3p}{2p_0}\,\frac{1}{p_0+1}\,\,\varphi_\uparrow^\dagger \varphi_\uparrow\,  +\int \frac{d^3p}{2p_0}\,\frac{1}{p_0-1}\,\,\varphi_\downarrow^\dagger \varphi_\downarrow
  \, .                                                      \label{s=1/2}
 \end{equation}
The momentum operators  are 
\begin{eqnarray}
&&  \hskip-1cm   i \Pi_{0}= \begin{pmatrix}
              i\Pi_{0,\uparrow} &0\\[4pt]
              0 &
              i\Pi_{0,\downarrow}
             \end{pmatrix} = \sqrt{\zeta \Lambda}\, 
\begin{pmatrix}
                      M_{04} & 0\\[4pt]
                      0 & - M_{04}
                     \end{pmatrix}                                 \nonumber   \\[12pt]
                &&    \hskip-1cm   i \Pi_{i}= \begin{pmatrix}
              i\Pi_{i,\uparrow} &0\\[4pt]
              0 & i\Pi_{i,\downarrow}
             \end{pmatrix} = \sqrt{\zeta \Lambda}\,\begin{pmatrix}
                      M_{0i}+M_{i4} & 0\\[4pt]
                      0 & M_{0i}- M_{i4}
                     \end{pmatrix} ,                                  \nonumber    \\[8pt]
\text{with}                                           \nonumber\\[8pt]
 &&  \hskip-1cm  M_{04}=\begin{pmatrix}
         (\rho - \frac{3i}{2}) \, p_0 -i p_0 p_i \frac{\p}{\p p_i} & \frac i2 \sigma_ip^i \\[8pt]
        \frac i2 \sigma_ip^i  & (\rho - \frac{3i}{2}) \, p_0 -i p_0 p_i \frac{\p}{\p p_i}
        \end{pmatrix}    ,       \nonumber
\\[16pt]
 &&    \hskip-1cm   M_{0i}+M_{i4}=          \nonumber \\[12pt]
   &&  \hskip-1cm 
 = \begin{pmatrix}
   (\rho - \frac{3i}{2}) \, p_i +i ( p_0+1)  \frac{\p}{\p p^i} -i p_i p_k \frac{\p}{\p p_k}+\frac 12\, \epsilon_{ijk}p^j\sigma^k 
 &- \frac i2 (p_0+1) \sigma_i \\[8pt]
 - \frac i2 (p_0+1) \sigma_i &
   (\rho - \frac{3i}{2}) \, p_i +i ( p_0+1)  \frac{\p}{\p p^i} -i p_i p_k \frac{\p}{\p p_k}+\frac 12\, \epsilon_{ijk}p^j\sigma^k 
 \end{pmatrix}   . \nonumber
\end{eqnarray}
\vskip10pt

We  first solve  equation (\ref{equ}) in the upper subspace, $\cH_\uparrow$. Since we are  dealing  with spinors, it takes more steps to come to the nontrivial part, to radial equation: some details  of the derivation are given in  Appendix~1. The  Laplacian has the form  
\begin{eqnarray}
 &&\hskip-1cm\Delta_\uparrow=\begin{pmatrix}
                       A_\uparrow  & B_\uparrow  \\[4pt] B_\uparrow  & A_\uparrow 
                      \end{pmatrix} ,   \quad \text{with}         \\[8pt]
 && \hskip-1cm  A_\uparrow  = \frac{15}{4} +\rho^2 +3i(\rho-2i)p_0 -i (p_0+1) \,\epsilon_{ijk} \, p^i \frac{\p}{\p p_j} \sigma^k  
 +\frac{p_0+1}{p_0-1}\,L_i L^i +\frac{2p_0^2}{p_0-1}\,\Big( p_i\frac{\p}{\p p_i}\Big)^2  \nonumber\\[4pt]
 && \hskip-1cm   \phantom{A_\uparrow  = }
 +\left( 2i\rho p_0 +2\, \frac{3p_0^2-p_0-1}{p_0-1}\right) p_i\frac{\p}{\p p_i}\ ,   \label{A} \\[4pt]
 &&   \hskip-1cm  B_\uparrow = -i (\rho - 2i) \sigma_ip^i +(p_0+1)^2\sigma_i\,\frac{\p}{\p p_i} - p_i \sigma^i\, p_k\frac{\p}{\p p_k}\ ,      \label{B}
\end{eqnarray}
 $\,L_i$ is the orbital part of the angular momentum, $\, L_i = i\epsilon_{ijk}\, p^j \p^k\, $, $\,\vec L^2=-L_iL^i$.
Introducing  (\ref{bispinors}) to  eigenvalue equation, after some transformations we obtain  
\begin{equation}
 \Delta_{\uparrow,eff}\, \varphi_\uparrow = - M^2 \varphi_\uparrow \, ,       \label{equphi}
\end{equation}
where the effective operator $ \Delta_{\uparrow,eff}\, $  is given by
\begin{eqnarray}
&& \Delta_{\uparrow,eff}\equiv \frac{1+p_0}{2}\left( A_\uparrow - \frac{p_k\sigma^k}{1+p_0}\, A_\uparrow \,\frac{p_i\sigma^i}{1+p_0}+
 \left[B_\uparrow, \frac{p_i\sigma^i}{1+p_0}\right]\right) \nonumber  \\
 && \phantom{ \Delta_{\uparrow,eff}}
 = \big(\rho+\frac i2\big)^2 +2( i\rho+1)\, p_0\, \frac{\p}{\p p}\, p + 2\,\frac{\p}{\p p}\, p +\frac{1+p_0}{1-p_0}\,\vec L^2 -\frac{2p_0^2}{1-p_0}\,p \,\frac{\p}{\p p}\, p \,\frac{\p}{\p p} \, . \qquad 
 \label{delta_up_eff}
\end{eqnarray}
Analogously, in the lower subspace $\cH_\downarrow$ we find
\begin{eqnarray}
&& \Delta_{\downarrow,eff}\equiv \frac{1-p_0}{2}\left( A_\downarrow - \frac{p_k\sigma^k}{1-p_0}\, A_\downarrow \,\frac{p_i\sigma^i}{1-p_0}-
 \left[B_\downarrow, \frac{p_i\sigma^i}{1-p_0}\right]\right) \nonumber  \\
 && \phantom{ \Delta_{\downarrow,eff}}
 =  \big(\rho+\frac i2\big)^2 - 2( i\rho+1)\, p_0\, \frac{\p}{\p p}\, p + 2\,\frac{\p}{\p p}\, p +\frac{1-p_0}{1+p_0}\,\vec L^2 -\frac{2p_0^2}{1+ p_0}\,p \,\frac{\p}{\p p}\, p \,\frac{\p}{\p p} \, . \qquad           \label{delta_down_eff}
\end{eqnarray}
The two effective  operators  differ by the change of sign, $\,p_0\to -p_0$, that is 
$\, \Delta_{\downarrow,eff}(p_0)=\Delta_{\uparrow,eff}(-p_0)\,$.

The $ \,\Delta_{\uparrow,eff}\,$ is spherically symmetric and has no terms that include  spin:  it commutes  with  $\,\vec J^2$, $J_3$ and $\vec L^2$, so we can label its solutions by the corresponding quantum numbers \,$j$, $m$ and $l$. Linearly independent solutions to (\ref{equphi}),  denoted by $\,\varphi_\uparrow\,$ and $ \chi_\uparrow$, can be written as
\begin{equation}
 \varphi_\uparrow(\vec p) = \frac{f_\uparrow(p)}{p}\, \varphi_{jm}(\theta,\varphi)\, ,
 \quad  j=l+\frac 12\, ,\ 
 \qquad  \chi_\uparrow(\vec p) = \frac{h_\uparrow(p)}{p}\, \chi_{jm}(\theta,\varphi)\, ,
 \quad j=l-\frac 12\, ,                                            \label{functions}
\end{equation}
where $\varphi_{jm}$ and $\chi_{jm}$ are the spin spherical harmonics, \cite{Bjorken}. Introducing the variable  $z$,
\begin{equation}
 z= \sqrt{\frac{p_0-1}{p_0+1}} \in(0,1) \, ,                 \label{z}
\end{equation}
from (\ref{equphi}) we find the radial equation for $\, f_\uparrow $,
\begin{eqnarray}
  (1-z^2)\,\frac{d^2  f_\uparrow }{dz^2} +( 2i\rho-1) z \, \frac{df_\uparrow}{dz}+ \left(\big(\rho +\frac i2\big)^2-M^2-\frac{l(l+1)}{z^2} \right) f_\uparrow =0  \, .         \label{equz}
\end{eqnarray}
The equation for $\, h_\uparrow \,$ is identical (with different $l$, (\ref{functions})), so  in the following we will discuss only  $\,f_\uparrow$.  Solutions to (\ref{equz}) are
\begin{eqnarray}
&& f_{\uparrow,1} =z^{-l} \, F\big(\frac 14 -\frac l2 -\frac{i\rho}{2}-\frac{iM}{2}\, , \frac 14 -\frac l2-\frac{i\rho}{2} +\frac{iM}{2}\, ; \frac 12-l\, ; z^2\big)   \,, \label{sol1}            \\[8pt]
&& f_{\uparrow,2} =z^{l+1}\, F\big(\frac 34 +\frac l2 -\frac{i\rho}{2}-\frac{iM}{2}\, , \frac 34 +\frac l2-\frac{i\rho}{2} +\frac{iM}{2}\, ; \frac 32+l\, ; z^2\big) \, ,                \label{sol2}
\end{eqnarray}
where $\, F(a,b,;c,z)\equiv\,  _2F_1(a,b,;c,z)\, $ is the hypergeometric fuction. As at  $z=0\,$ first solution is divergent and  second  finite,  $ \,f_{\uparrow,2}\,$ is the physical one.

Equation for $\,\Delta_\downarrow\,$ in $\cH_\downarrow$ is completely analogous. For the radial part we obtain
\begin{eqnarray}
  (1-w^2)\,\frac{d^2  f_\downarrow }{dw^2} + (2i\rho -1) w \, \frac{df_\downarrow}{dw}+ \left(\big(\rho +\frac i2\big)^2-M^2-\frac{l(l+1)}{w^2} \right) f_\downarrow =0  \, ,        \label{equw}
\end{eqnarray}
that is, an equation identical to (\ref{equz}) only written in variable $w\,$,
\begin{equation}
 w= \sqrt{\frac{p_0+1}{p_0-1}}= \frac 1z \in(1,\infty) .
\end{equation}
Therefore  we have 
\begin{eqnarray}
&& f_{\downarrow,1} =w^{-l}\, F\big(\frac 14 -\frac l2-\frac{i\rho}{2} -\frac{iM}{2}\, , \frac 14 -\frac l2 -\frac{i\rho}{2}+\frac{iM}{2}\, ; \frac 12-l\, ; w^2\big)  \, , \\[8pt]
&& f_{\downarrow,2} =w^{l+1}\, F\big(\frac 34 +\frac l2 -\frac{i\rho}{2}-\frac{iM}{2}\, , \frac 34 +\frac l2-\frac{i\rho}{2} +\frac{iM}{2}\, ; \frac 32+l\, ; w^2\big) \, .    \label{sol22}
\end{eqnarray}
Analyzing the asymptotics of these functions for $w\to\infty$, we find that they behave in the same way,
\begin{equation}
 f_{\downarrow,1}(w),\, f_{\downarrow,2}(w) \Big|_{w\to\infty}\sim  w^{i\rho-\frac 12} \,\big(Cw^{iM} +C^* w^{-iM}\big) .                  \label{M^2}
\end{equation}
For real $M$, both solutions are asymptotically combinations of plane waves in  $\,\log w$ (up to a multiplicative function); for $M^2<0\,$, solutions diverge.

This asymptotics eventually results in the $\delta$-function normalisation of the solutions. To show that, let us consider  the scalar product in more details. Calculating the  product of two functions of the form (\ref{functions}) we obtain
\begin{eqnarray}
 &&(\psi_\uparrow,\psi_\uparrow') = \int \frac{d^3p}{ 2 p_0 (1+p_0)}\, \varphi_\uparrow^\dagger \varphi'_\uparrow =\delta_{jj'} \delta_{mm'}\int\limits_0^1 dz\, (f_\uparrow^* f_\uparrow' + h_\uparrow^* h_\uparrow' )\ ,                   \label{70} \\
 && (\psi_\downarrow,\psi_\downarrow') = \int \frac{d^3p}{2 p_0 (1-p_0)}\, \varphi_\downarrow^\dagger \varphi'_\downarrow = - \delta_{jj'} \delta_{mm'}\int\limits_1^\infty dw\,  ( f_\downarrow^* f_\downarrow' +  h_\downarrow^* h_\downarrow')\ .  
 \end{eqnarray}
Therefore taking into account (\ref{(s=1/2)}) we have
\begin{eqnarray}
&& (\Psi,\Psi') = 
 \delta_{jj'} \delta_{mm'}\int\limits_0^1 dz\, \left(f^*_\uparrow f'_\uparrow+ h^*_\uparrow h'_\uparrow\right)   +
 \delta_{jj'} \delta_{mm'}\int\limits_1^\infty dw\, \left(f^*_\downarrow f'_\downarrow+ h^*_\downarrow h'_\downarrow\right)  \label{pro}             \\[4pt]
 && \phantom{ (\Psi,\Psi')}
 =\delta_{jj'} \delta_{mm'}\int\limits_0^\infty dz\, (f^* f'+ h^*h') \    .        \label{product}
\end{eqnarray}
In the last line we  extended the radial functions  $\,f_{\uparrow, \downarrow}(z)$, $h_{\uparrow, \downarrow}(z)\,$ to   complete  semiaxis $\, z>0\,$  by gluing them smoothly at $\, z=1$:
\begin{equation}
 f(z) =\left\{ \begin{array}{ll}
                f_\uparrow(z),\quad & z\in(0,1)\\[6pt]
                  f_\downarrow(z),\quad & z\in(1,\infty)
               \end{array}
  \right. , \qquad
  \ h(z) =\left\{ \begin{array}{ll}
                h_\uparrow(z),\quad & z\in(0,1)\\[6pt]
                  h_\downarrow(z),\quad & z\in(1,\infty)
               \end{array}   
  \right.    .
\end{equation}

Hence, the physical solution of  (\ref{equw}) is the  smooth continuation of (\ref{sol2}) over  $z=1$, that is $\, f_{2,\downarrow}$. 

In summary, the eigenfunctions of the Laplace operator $\,\Delta$ are given by
\begin{eqnarray}
  && \hskip-10pt\varphi_{Mjm}(z,\theta,\varphi) = C (1-z^2)\, z^{l}\, F\big(\frac 34 +\frac l2 -\frac{i\rho}{2}-\frac{iM}{2}\, , \frac 34 +\frac l2-\frac{i\rho}{2} +\frac{iM}{2}\, ; \frac 32+l\, ; z^2\big) \, \varphi_{jm}(\theta,\varphi)  \label{a} 
  \\[8pt]
&&\phantom{ \hskip-10pt\varphi_{Mjm}(z,\theta,\varphi)   }
= C \,(1-z^2)\, z^{j-\frac 12}\, F\Big(\frac{j+1-i\rho-iM}{2}\,, \frac{j+1-i\rho+iM}{2}\,;j+1,z^2\Big)\, \varphi_{jm}(\theta,\varphi)      \nonumber 
 ,                \\[12pt]
 && \hskip-10pt\chi_{Mjm}(z,\theta,\varphi) 
 = C'(1-z^2)\, z^{l}\, F\big(\frac 34 +\frac l2 -\frac{i\rho}{2}-\frac{iM}{2}\, , \frac 34 +\frac l2 -\frac{i\rho}{2}+\frac{iM}{2}\, ; \frac 32+l\, ; z^2\big) \, \chi_{jm}(\theta,\varphi) \label{b} \\[8pt]
 &&  \phantom{ \hskip-10pt\chi_{Mjm}(z,\theta,\varphi)}
 =C' \,(1-z^2)\, z^{j+\frac 12}\, F\Big(\frac{j+2-i\rho-iM}{2}\,, \frac{j+2-i\rho+iM}{2}\,;j+2,z^2\Big)\, \chi_{jm}(\theta,\varphi), \qquad\qquad   \nonumber  
\end{eqnarray}
with  $z\in(0,\infty)\,$. The  spectrum of the Laplacian is continuous, $M^2\in(0,\infty)$: the mass parameter $M$ can be taken to be positive, $\, M\geq0$, as $\, \Psi_{M\,jm} =\Psi_{-M\,jm}  $. Each value of $M\,$ is infinitely degenerate, with degeneracy  $\, 2\times\sum_j (2j+1)$, $\, j = \frac 12, \frac 32,\dots$.  Normalization of the eigenfunctions is essentially determined by their asymptotics  at $\,z\to\infty\,$, which implies
\begin{equation}
 (\varphi_{Mjm},\varphi_{M'j'm'} )\sim  \delta_{jj'}\,\delta_{mm'}\,\delta(M-M')\, .
\end{equation}
 Completeness and orthogonality of the radial functions can be shown exactly, see Appendix~2.

The trick (that is, the method) of writing the sum of two scalar products in different subspaces as one integral provides in fact a proof of hermiticity for  spherically-symmetric operators on $\,\cH_\uparrow\oplus\cH_\downarrow\,$. Let us outline the  proof: to be concrete, we consider the Laplace operator, but the reasoning is very similar in other cases. Assume that we have two radial functions of the type (\ref{functions}) with
\begin{equation}
  \varphi(p) = \frac{f(p)}{p}\, , \qquad\varphi'(p) = \frac{g(p)}{p} \ .
\end{equation}
The matrix elements of the Laplacian are defined by  (\ref{70}-\ref{pro}). We find that in the radial subspace
\begin{equation}
 (\varphi, \Delta_\uparrow \,\varphi') -(\Delta_\uparrow \, \varphi  ,\varphi'  )=
  2i\rho zf^*g\Big\vert_0^1 +(1-z^2)\Big(f^*\, \frac{dg}{dz}+ \frac{df^*}{dz} \,g\Big) \Big\vert_0^1\ .
  \label{integral}
\end{equation}
Evaluating this expression for the eigenfunctions (\ref{a}-\ref{b}) we can verify that  the boundary contribution at $z=1\,$  does not vanish. This means that $\,\Delta_\uparrow\,$ is only {\it formally} self-adjoint, that is, the difference  $\, \big(\varphi, (\Delta_\uparrow -\Delta_\uparrow^\dagger)\varphi'\big)\, $ is not always zero.  To achieve self-adjointness,  additional boundary conditions at $ z=1\,$   have to be imposed: they restrict the initial  $\,\cH_\uparrow\,$ in such way that  (\ref{integral}) vanishes in all states, \cite{Pim}.  A more detailed analysis shows that the appropriate boundary conditions in fact imply discreteness of the spectrum of  $\,\Delta_\uparrow$.

The matrix elements of $\,\Delta\,$  defined on $\,\cH_\uparrow\oplus\cH_\downarrow\,$  are given by  exactly  same expression (\ref{integral}), except that the boundary terms are  evaluated at $ 0$ and $\, \infty$. The reason is that $\,\Delta_\downarrow$ is obtained from $\,\Delta_\uparrow$  by replacement $\, p_\mu\to -p_\mu$, that is (for spherically symmetric operators) by  $\,z\to w$; otherwise the form of the operator is exactly the same. Therefore, when evaluating  (\ref{integral})  at   $\ \Big\vert_0^\infty\ $ there will be no special conditions at $\, z=1$, or more precisely, the  boundary terms for $\,\Delta_\uparrow\,$ at $z=1-\epsilon\,$ will cancel the boundary terms of $\,\Delta_\downarrow\,$ at $z=1+\epsilon\,$, when evaluated for continuous functions with continuous first derivative at $z=1$. One can check that  conditions at $\,z=\infty\,$ coming from (\ref{integral}) do not impose further restrictions compared to those that come from normalization.

For comparison and future reference, we add the results for $\,(\rho,s=0)$. The Laplacian  is
\begin{equation}
 \Delta_{\uparrow}
 =\rho^2 +\frac 14 +\frac 12 \big(1+ 2 i\rho\big) p_0 +  \big(2 i\rho p_0 + 4p_0 +2\big)\,\frac{\p}{\p p}\, p +\frac{1+p_0}{1-p_0}\,\vec L^2 -\frac{2p_0^2}{1-p_0}\,p \,\frac{\p}{\p p} \,\frac{\p}{\p p} \,p  \, ,
\end{equation}
and  the eigenfunctions are of the form
\begin{equation}
 \psi_{M l m,\uparrow}(\vec p) = \frac{f_{\uparrow}(p)}{p}\, Y_l^m(\theta, \varphi)\, .
\end{equation}
We obtain the radial equation 
\begin{eqnarray}
  (1-z^2)\,\frac{d^2  f_\uparrow }{dz^2} + 2i\rho z \, \frac{df_\uparrow}{dz}+ \left(\big(\rho -\frac i2\big)^2-M^2-\frac{l(l+1)}{z^2}+\frac{1+2i\rho}{1-z^2} \right) f_\uparrow =0  \,   ,       \label{equz0}
\end{eqnarray}
with solutions that are almost identical to (\ref{sol1}-\ref{sol2}),
\begin{eqnarray}
&& f_{\uparrow,1} =z^{-l} \, (1-z^2)^{\frac{1-i\rho}{2}}\,\, F\big(\frac 14 -\frac l2 -\frac{i\rho}{2}-\frac{iM}{2}\, , \frac 14 -\frac l2 -\frac{i\rho}{2}+\frac{iM}{2}\, ; \frac 12-l\, ; z^2\big) \\[8pt]
&& f_{\uparrow,2} =z^{l+1}\, (1-z^2)^{\frac{1-i\rho}{2}}\, \, F\big(\frac 34 +\frac l2 -\frac{i\rho}{2}-\frac{iM}{2}\, , \frac 34 +\frac l2 -\frac{i\rho}{2}+\frac{iM}{2}\, ; \frac 32+l\, ; z^2\big) \, .
\label{c}
\end{eqnarray}
As before, the second one is physical. Extension to  $\cH_\downarrow$ and properties of $\,\Delta_\downarrow$  are  analogous to $s=\frac 12$:  complete solution is obtained by continuation of $\, f_{\uparrow,2}\,$  to the interval $z\in(0,\infty)$. 

We will in Appendix~2 use the fact that this solution is real, so let us prove it. Using the properties of  hypergeometric function \cite{Abramowitz}, we find
\begin{eqnarray}
 && F\big(\frac 34 +\frac l2 -\frac{i\rho}{2}-\frac{iM}{2}\, , \frac 34 +\frac l2 -\frac{i\rho}{2}+\frac{iM}{2}\, ; \frac 32+l\, ; z^2\big)  =    \nonumber \\[4pt]
 && \qquad\qquad  = (1-z^2)^{i\rho}\, 
 F\big(\frac 34 +\frac l2 +\frac{i\rho}{2}-\frac{iM}{2}\, , \frac 34 +\frac l2 +\frac{i\rho}{2}+\frac{iM}{2}\, ; \frac 32+l\, ; z^2\big)      \, ,  \nonumber
\end{eqnarray}
that is, 
\begin{equation}
f_{2,\uparrow} =z^{l+1}\,\sqrt{1-z^2}\, (1-z^2)^{-\frac{i\rho}{2}}\,  F\big(\frac 34 +\frac l2 -\frac{i\rho}{2}-\frac{iM}{2}\, , \frac 34 +\frac l2 -\frac{i\rho}{2}+\frac{iM}{2}\, ; \frac 32+l\, ; z^2\big)          =      f^*_{2,\uparrow}\ .   \label{84}
\end{equation}

 The spectrum of $\,\Delta\,$ for $s=0$ is, up to the spin degeneracy,  the same as the spectrum of $\,\Delta\,$ for $s=\frac 12$.

\section{Discussion}

To the comments and discussion given in the text already we  add the following.

The main idea in this paper was to find the Laplace operator on fuzzy de~Sitter space and study its properties. The immediate motivation was a possible application to cosmology and inflation, but  the overall idea is more general: to understand whether the behavior of matter on  noncommutative spaces can  provide some explanations of  semiclassical or quantum effects in gravity. In the commutative case,  Laplacian is one of the main  geometric characteristics of a manifold; it is in the noncommutative frame formalism defined in a  systematic way as well. Concrete expression for the Laplacian depends on  degree of  differential form which it acts upon: here, in addition, we consider the action of the Laplacian on wave functions that define the representation space  $\,\cH$ (the `quantum-mechanical' Laplacian). In fact, the main results of the paper concern noncommutative quantum mechanics on fuzzy de Sitter space; properties of the scalar field will be discussed separately. 

We obtained that the fuzzy de~Sitter Laplacian, defined straightforwardly, is a non-hermitian operator, both when acting on wave functions $\Psi$, (\ref{DeltaPsi}) and on scalar fields $f$, (\ref{Deltaf}). The meaning of this result is not quite clear. It might be an implication of the (unnecessary) generality of the frame formalism, which (among other assumptions) postulates that the frame derivations $e_a$ obey the `reality condition', that is that momenta $\Pi_a$ are defined as antihermitian operators. On the other hand, this assumption seems to be quite logical and works very well in a number of examples. The commutative Laplacian is always hermitian:  we therefore symmetrized (\ref{DeltaPsi}) to obtain a hermitian expression. We then solved the corresponding quantum-mechanical equation and found that the spectrum of the Laplacian is continuous, given by eigenvalue $\,M\in(0,\infty)$, and that the  eigenfunctions behave  as spherical waves. The eigenfunctions $\,\Psi_{Mjm}\,$ are given by (\ref{a}-\ref{b}) and (\ref{c}) for, respectively,  $(\rho,s=\frac 12)$ and $(\rho,s=0)$ Moylan representations; each eigenvalue has a degeneracy in the angular momentum.  In comparison with the analogous result for commutative de~Sitter space we see the (expected) reduction of the number of degrees of freedom: the corresponding commutative solutions, along with $M$, $j$ and $m$, have the eigenvalue of energy $\omega$ as a quantum number, \cite{BD}. It is, however, interesting to observe that radial functions which appear in  commutative and noncommutative solutions can  both be expressed as  Jacobi functions, with different values of parameters $\alpha$ and $ \beta$, \cite{Koornwinder}. 

It is possible  to solve the eigenvalue equation for the non-hermitian Laplacian  (\ref{deltapsi}) as well. For $s=0\,$, the operator $\,\triangle\,$ is in $\,\cH_\uparrow\,$ given by
\begin{equation}
 \triangle_{\uparrow}
 =\rho^2 +\frac 14- \big(1+ 2 i\rho\big) p_0 +  \big(2 i\rho p_0 + p_0 +2\big)\,\frac{\p}{\p p}\, p +\frac{1+p_0}{1-p_0}\,\vec L^2 -\frac{2p_0^2}{1-p_0}\,p \,\frac{\p}{\p p} \,\frac{\p}{\p p} \,p  \, .
\end{equation}
The radial part of the corresponding eigenvalue equation is
\begin{eqnarray}
  (1-z^2)\,\frac{d^2  f_\uparrow }{dz^2} + ( 2i\rho-3) z \, \frac{df_\uparrow}{dz}+ \left(\big(\rho +i \big)^2+\frac 94 -M^2-\frac{l(l+1)}{z^2}-\frac{2(1+2i\rho)}{1-z^2} \right) f_\uparrow =0  \,       
\end{eqnarray}
and has solutions
\begin{eqnarray}
&& f_{\downarrow,2} =\frac{z^{-l}}{1-z^2}\, \, F\big(\frac l2 -\frac{i\rho}{2}-\frac i2\, \sqrt{M^2-\frac 94}\, ,  \frac l2 -\frac{i\rho}{2}+\frac i2\, \sqrt{M^2-\frac 94}\,; \frac 32+l\, ; z^2\big)
\\[4pt]
&& f_{\uparrow,2} =\frac{z^{l+1}}{1-z^2}\, \, F\big(\frac l2 -\frac{i\rho}{2}-\frac i2\, \sqrt{M^2-\frac 94}\, ,  \frac l2 -\frac{i\rho}{2}+\frac i2\, \sqrt{M^2-\frac 94}\,; \frac 32+l\, ; z^2\big) \, .
\end{eqnarray}
The solutions are of course not normalizable: they all diverge at $z=1\,$. 

Still, it is interesting and perhaps important that the Laplacian on the fuzzy de Sitter space $\triangle\,$ obtained from `the first principles' that is, directly from definitions, is non-hermitian. It gives a model of non-unitary evolution in the gravitational field (which Hawking considered as possible solution to the information paradox). Indeed, antihermitian part of the Laplacian, $\,V=-3\sqrt{\zeta\Lambda}\,\Pi_0\, $, implies non-conservation of probability; non-hermitian terms have been used before, in nuclear physics, to describe the $\alpha$-decay. It would be interesting to see whether the transition amplitudes $\, \langle \Psi_{Mjm}\vert V\vert \Psi_{M'jm}\rangle\, $, calculated in perturbation theory, could in any way be  related to the Hawking flux. 

If we stick to  conservative interpretation and to hermitian ordering, there is another  interesting problem that one can pose in the given framework: to solve the quantum-mechanical  equation for the scalar particle in the potential of  inflatory type, and compare the results with the effects obtained in the standard de~Sitter cosmology.

 An important problem which needs further investigation is to solve classical equation of motion (\ref{EQ}),  and then find the propagator of the noncommutative scalar field $f$. It would seem that this problem is not easily reducible to the pure representation theory. The scalar field can be expanded  as
\begin{equation}
 f=\sum_{M,j,m,M',j',m'} c_{jm,j'm'}\vert\Psi_{Mjm}\rangle\langle\Psi_{Mj'm'}\vert 
\end{equation}
and (\ref{EQ}) becomes an equation for the coefficients $\, c_{Mjm,M'j'm'}\,$.  As
\begin{equation}
  [\Pi_\mu,[\Pi^\mu,f]]  = \Pi_\mu\Pi^\mu\,  f + f\, \Pi_\mu\Pi^\mu -2\Pi_\mu f\Pi^\mu \, ,
\end{equation}
(denoting the mass of the scalar field by ${\rm M}$) we have 
\begin{equation}
  \sum ({\rm M}^2+M^2 +{M'}^2)\,
 c_{Mjm,M'j'm'}\vert\Psi_{Mjm}\rangle\langle\Psi_{Mj'm'}\vert   = 2 \sum c_{jm,j'm'}\, \Pi_\mu\, \vert\Psi_{Mjm}\rangle\langle\Psi_{Mj'm'}\vert\, \Pi^\mu  \ .
\end{equation}
Since the  action of the Laplacian is off-diagonal, this equation is (in principle) complicated; however, even if not fully solved, it might  provide an insight into the structure of the scalar field propagator.

\vskip0.8cm
\begin{large}
{\bf Acknowledgement}
\end{large}
 This work was supported by the Serbian 
Ministry of Education, Science and Technological Development Grant 451-03-9/2021-14/200162.

\vskip1cm

\begin{Large}
{\bf Appendix  1}
\end{Large}

In this appendix we give details of how to solve (\ref{equ})
for  $s=\frac 12$. In the signature that we use we have
\begin{align}
    &\vec p=(p_i), \qquad \vec r=(x^i)=i\nabla^i=i\frac{\partial}{\partial p_i}, \qquad \vec L=(L_i)=\vec r \times \vec p, \qquad L_i=i\epsilon_{ijk}p^j \frac{\partial}{\partial p_k}, \nonumber \\[4pt]
    &\vec \sigma=(\sigma_i), \qquad \sigma_i\sigma_j=-\eta_{ij}-i\epsilon_{ijk}\sigma^k, \qquad \epsilon_{ijk}\epsilon^{imn}=-(\delta^m_j \delta^n_k-\delta^n_j \delta^m_k), \qquad \epsilon_{ijk}\epsilon^{ijn}=-2\delta^n_k  \ .
\end{align}
In  $\cH_\uparrow\,$ the Laplacian has the form $\Delta_\uparrow=\begin{pmatrix}
                       A_\uparrow  & B_\uparrow  \\[4pt] B_\uparrow  & A_\uparrow
                      \end{pmatrix} \, $ 
and the scalar product  is given by (\ref{(s=1/2)}), (\ref{s=1/2}). Applying  $\,\Delta_\uparrow\,$ to  bispinor (\ref{bispinors}) in the eigenvalue equation,  we obtain 
 \begin{eqnarray}
  && A_\uparrow \varphi_\uparrow +B_\uparrow\,
  \frac{p_k\sigma^k}{1+p_0}\, \varphi_\uparrow = - M^2\varphi_\uparrow\\[4pt]
  && B_\uparrow \varphi_\uparrow +   \frac{p_k\sigma^k}{1+p_0}\,  A_\uparrow\,
  \frac{p_k\sigma^k}{1+p_0}\, \varphi_\uparrow = - M^2 \frac{p_k\sigma^k}{1+p_0}\, \varphi_\uparrow\ .
 \end{eqnarray}
After some transformations, introducing
 \begin{equation}
   \Delta_{\uparrow,eff}\equiv \frac{1+p_0}{2}\left( A_\uparrow - \frac{p_k\sigma^k}{1+p_0}\, A_\uparrow \,\frac{p_i\sigma^i}{1+p_0}+
 \left[B_\uparrow, \frac{p_i\sigma^i}{1+p_0}\right]\right)
\end{equation}
 these two equations can be reduced to one
\begin{equation}
 \Delta_{\uparrow, eff}\, \varphi_\uparrow= - M^2 \varphi_\uparrow  .
\end{equation}
Using  $\, A_\uparrow$ and $\, B_\uparrow$ (\ref{A}-\ref{B}), we arrive at 
\begin{align}
    \Delta_{\uparrow,eff}
 = \big(\rho+\frac i2\big)^2 +2( i\rho+1)\, p_0\, \big(1+p_i \,\frac{\p}{\p p_i}\big)\, + 2\,\big(1+p_i \,\frac{\p}{\p p_i}\big) +\frac{1+p_0}{1-p_0}\,\vec L^2 -\frac{2p_0^2}{1-p_0} \,p_i \,\frac{\p}{\p p_i}\, \big(1+p_i \,\frac{\p}{\p p_i}\big) \, ,                 \nonumber
\end{align}
and by passing to the spherical coordinate system, the effective Laplacian adopts the form  (\ref{delta_up_eff}) given in the text.

\vskip1cm

\begin{Large}
{\bf Appendix 2}
\end{Large} 

In this appendix we outline the proof of orthogonality and completeness of  eigenfunctions of the Laplacian: as simpler, we discuss  eigenfunctions in the $(\rho, s=0)$ representations
\begin{eqnarray}
 && \Psi_{Mlm}(\vec p)=C_{Ml}\, \,\frac{f_{Ml}(p)}{p}\, Y_l^m(\theta,\varphi)              \label{star} \\[4pt]
 && \phantom{ \Psi_{Mlm}(\vec p)}  =\frac {C_{Ml}}{2}\, z^l\,(1-z^2)^{\frac{3-i\rho}{2}}\, 
  F\big(\frac 34 +\frac l2 -\frac{iM}{2}-\frac{i\rho}{2}\, , \frac 34 +\frac l2 +\frac{iM}{2}-\frac{i\rho}{2}\, ; \frac 32+l\, ; z^2\big)\, Y_l^m \, .\qquad
 \label{starstar}
\end{eqnarray}
The scalar product of two functions of the form (\ref{star}) is given by
\begin{equation}
 (\Psi, \Psi') = C^* C'\,\delta_{ll'}\, \delta_{mm'} \int\limits_0^\infty \frac{dz}{\vert 1-z^2\vert}\, f^*_{Ml}\,  f'_{M'l'} \ .
\end{equation}
Orthogonality can be shown by expressing the hypergeometric function in (\ref{starstar}) as the Jacobi function, and using the properties of the Jacobi transform and its inverse. In the following, we use the notation of \cite{Koornwinder}  and the results presented there. The Jacobi functions are defined as
\begin{equation}
 \phi^{(\alpha,\beta)}_\lambda (t) = F\Big(\, \frac 12 \,(\alpha+\beta+1-i\lambda )\,, \frac 12\, (\alpha+\beta+1-i\lambda )\, ;\, \alpha+1; -\sinh^2t\Big)\, ,
\end{equation}
 for $\alpha, \, \beta,\, \lambda\in \mathbb{C}$, $\, \alpha \neq -1,-2,\dots$ .
 Therefore radial eigenfunctions entering (\ref{star}) can be rewritten as
 \begin{equation}
  f_{Ml} =  f_{Ml}^* = C_{Ml}^* \, z^{l+1}\,\sqrt{\vert 1-z^2\vert }\, (1-z^2)^{\frac{i\rho}{2}}\, \,
  \phi^{(\alpha,\beta)}_\lambda (t)  \, ,                         \label{101}
 \end{equation}
with 
\begin{equation}
  \alpha = l+\frac 12\, , \quad \beta = i\rho, \quad \lambda = M, \quad  i\sinh t = z . \label{parameters}
\end{equation}
Clearly, integration in formulas for the Jacobi transform will in our case be done along the imaginary axis $\, (-i\infty, \, i\infty)\,$, using the analytic continuation of the hypergeometric function. A rigorous derivation of the necessary steps is beyond the scope of this text;  we shall use the formulas as given in \cite{Koornwinder}, where they are proved for real $\,\beta$, $\alpha > -1$, and then analitically continued to complex values of  parameters  $\,\beta,\alpha \in \mathbb{C}$, $\,\alpha \neq -1,-2,\dots$ . We start by referring to some definitions and theorems.
 
For $\alpha>-1$, $\,\vert\beta\vert<\alpha+1$, the Jacobi function is the kernel of the Jacobi transformation
\begin{eqnarray}
  \hat f(\lambda) =\int\limits_0^\infty dt\, (2\sinh t)^{2\alpha +1} (2\cosh t)^{2\beta+1}\, \phi^{(\alpha,\beta)}_\lambda(t)\, f(t)         \, .       \label{transform} 
  \end{eqnarray}
The inverse transformation is given by
\begin{eqnarray}
 f(t) =\frac{1}{2\pi} \int\limits_0^\infty d\lambda\,\, \vert c_{\alpha,\beta}(\lambda)\vert^{-2}\, \phi^{(\alpha,\beta)}_\lambda(t) \,\hat f(\lambda)           \, ,       \label{completeness}
\end{eqnarray}
where the constants $\, c_{\alpha,\beta}(\lambda)\,$ are\footnote{Jacobi functions are even, and the Jacobi transformation is a generalization of the Fourier cosine transformation, as $\,\phi^{(-1/2,-1/2)}_\lambda(t)  =\cos \lambda t\,$.  Therefore the integrals in  (\ref{transform}-\ref{completeness}) can be extended to  interval $(-\infty,\infty)$, assuming that $f$, $\hat f$ are even functions.}
\begin{equation}
 c_{\alpha,\beta}(\lambda) = \frac{\,2^{\alpha+\beta+1-i\lambda}\, \Gamma(\alpha+1)\, \Gamma(i\lambda)\,}{\, \Gamma\big( \frac 12\, (\alpha+\beta+1+i\lambda)\big)\,\Gamma\big( \,\frac 12 (\alpha-\beta+1+i\lambda)\big)\, }  \ .
\end{equation}
Equation (\ref{completeness}) holds when  $\, c_{\alpha,\beta}(\lambda)\, $, $\, c^{-1}_{\alpha,\beta}(\lambda) \,$ are finite i.e. have no poles, which is true in our  case, (\ref{parameters}). Applying the Jacobi transformation to Jacobi functions we obtain that they are continuously orthogonal,
\begin{equation}
\int\limits_0^\infty dt\, (2\sinh t)^{2\alpha +1} (2\cosh t)^{2\beta+1}\, \phi^{(\alpha,\beta)}_\lambda(t)\, \phi^{(\alpha,\beta)}_{\lambda'}(t) =
2\pi \, \vert c_{\alpha,\beta}(\lambda)\vert^{2}\, \delta(\lambda-\lambda')\, .
\end{equation}
The last formula gives in fact orthogonality of the eigenfunctions of the Laplacian. Indeed, since the eigenfunctions are real, we have
\begin{eqnarray}
 && \int\limits_0^\infty \frac{dz}{\vert 1-z^2\vert}\, f^*_{Ml}\,  f_{M'l} =
 \int\limits_0^\infty \frac{dz}{\vert 1-z^2\vert}\, 
 z^{2l+2}\, \vert 1-z^2\vert\, (1-z^2)^{{i\rho}}\, \phi^{(\alpha,\beta)}_M\, \phi^{(\alpha,\beta)}_{M'}\,\\
 && \phantom{} 
 \quad = \int\limits_0^{i\infty} {dt}\, 
 (\sinh t)^{2\alpha+1}\,  (\cosh t)^{2\beta+1}\, \phi^{(\alpha,\beta)}_{M}\,\phi^{(\alpha,\beta)}_{M'}\,
\end{eqnarray}
for values of parameters given by (\ref{parameters}). Therefore a possible choice of  normalization constants
$\, C_{Ml}\, $,
\begin{equation}
 \vert C_{Ml}\vert  = 2^{-l-1} \, \sqrt\pi \,\, \vert c_{\alpha,\beta}(M) \vert \, 
\end{equation}
is 
\begin{equation}
  C_{Ml}  = \frac{\,\sqrt{2\pi}\, \,\Gamma\big(l+\frac 32\big)\, \Gamma\big(iM\big)\,}{\, \Gamma\Big( \frac 12\, (l+\frac 32 +i\rho+iM)\Big)\,\Gamma\Big(  \frac 12\, (l+\frac 32 -i\rho+iM)\, \Big) } \ .
\end{equation}
Completeness of the set of functions (\ref{101}) is basically the formula for the inverse Jacobi transformation (\ref{completeness}).

%
%
 
 %
%


\begin{thebibliography}{99}


\bibitem{book}
  J.~Madore,
  ``An Introduction To Noncommutative Differential Geometry And Its Physical
  Applications'',
  Lond.\ Math.\ Soc.\ Lect.\ Note Ser.\  {\bf 257} (2000).



\bibitem{Buric:2015wta}
  M.~Buric and J.~Madore,
  Eur.\ Phys.\ J.\ C {\bf 75} (2015) no.10,  502
  [arXiv:1508.06058 [hep-th]].


\bibitem{Buric:2017yes}
  M.~Buric, D.~Latas and L.~Nenadovic,
  Eur.\ Phys.\ J.\ C {\bf 78} (2018) no.11,  953
  [arXiv:1709.05158 [hep-th]].

\bibitem{Buric:2019yau}
M.~Buric and D.~Latas,
Phys. Rev. D \textbf{100} (2019) no.2, 024053
[arXiv:1903.08378 [hep-th]].
  

\bibitem{fs}
  J.~Madore,
  Class.\ Quant.\ Grav.\  {\bf 9} (1992) 69.

\bibitem{Jurman:2013ota}
D.~Jurman and H.~Steinacker,
JHEP \textbf{01} (2014), 100
doi:10.1007/JHEP01(2014)100
[arXiv:1309.1598 [hep-th]].

  
\bibitem{BD}
N.D. Birrell and P.C.W. Davies, ``Quantum Fields in Curved Space'',
 Cambridge Monographs on Mathematical Physics (1982).

%
%
%

\bibitem{Moylan}
  P.~Moylan,
  J.\ Math.\ Phys.\  {\bf 24} (1983) 2706.


\bibitem{Bargmann}
  V.~Bargmann and E.~P.~Wigner,
  Proc.\ Nat.\ Acad.\ Sci.\  {\bf 34} (1948) 211.


\bibitem{Dobrev}
V.~K.~Dobrev, G.~Mack, V.~B.~Petkova, S.~G.~Petrova and I.~T.~Todorov,
``Harmonic Analysis  on the n-Dimensional Lorentz Group and Its Application to Conformal Quantum Field Theory'',
Lecture Notes in Physics, Springer (1977).


\bibitem{skripta}
D.~Skinner, ``Mathematical Methods'',
http://www.damtp.cam.ac.uk/user/dbs26/1Bmethods.html


\bibitem{Buric:2010xs}
M.~Buric, H.~Grosse and J.~Madore,
JHEP \textbf{07} (2010), 010
[arXiv:1003.2284 [hep-th]].
%


\bibitem{Madore_1999}
J.~Madore and H.~Steinacker,
Jour. Phys. A: Mathematical and General, {\bf 33} (1999) 327.
%


\bibitem{Cho_1999}
S.~Cho,
Jour. Phys. A: Mathematical and General, {\bf 32} (1999) 2091.



\bibitem{Wald}
R.~M.~Wald,
``Quantum Field Theory in Curved Space-Time and Black Hole Thermodynamics,''
Chicago Lectures in Physics (1995).


\bibitem{Bjorken}  
J.~D.~Bjorken, S.~D.~.~Drell, ``Relativistic Quantum Mechanics'',
McGraw-Hill (1965).



\bibitem{Pim}
V. Hutson, J. Pym, M. J. Cloud,
``Applications of Functional Analysis and Operator Theory'',
 Elsevier Science (2005)

 \bibitem{Abramowitz}
 M. Abramowitz, I. A. Stegun, ``Handbook of Mathematical Functions: With Formulas, Graphs, and Mathematical Tables'',  Dover Publications (1965).


 




\bibitem{Koornwinder}
T.~H.~Koornwinder, ``Jacobi Functions and Analysis on Noncompact Semisimple Lie Groups'' in 
``Special Functions: Group Theoretical Aspects and Applications'',
Springer (1984).

%
\end{thebibliography}
\end{document}